\documentclass[a4paper,12pt]{article}
\usepackage{amssymb}
\date{}
\title{Path Integral Approach to Fermionic Vacuum Energy in Non-parallel D1-Branes}
\author{A. Jahan\\
Department of Physics, Amirkabir University of Technology, Tehran, Iran\\jahan@aut.ac.ir}
\begin{document}
\maketitle
\begin{abstract}
The fermionic one loop vacuum energy of the superstring theory in a system of non-parallel D1-branes is derived by applying the path integral formalism.
\end{abstract}
\section*{\large 1\quad Introduction}
The present work is sequel to the previous one in which we applied the path integral technique to derive the one loop vacuum energy (zero point energy) of a bosonic string in a system of non-parallel D1-branes [1]. Here we shall derive the fermionic zero point energy by engaging the path integral formalism for a superstring in the same system of D1-branes. The path integral approach to the superstring theory is not a new subject. Indeed it stands as an alternative approach to unveil the physics of string theory through calculating the superstring S-matrix elements [2-9]. So, after a quick review of the path integral derivation of the fermionic partition function of the open superstrings,  a similar approach is followed to derive the fermionic partition function in a system of angled D1-branes in section 3. We show that the result is in agreement with one derived earlier using the harmonic oscillator representation [10-12].\\
\section*{\large 2\quad Fermionic Partition Function: Path Integral Approach}
From the fermionic part of the superstring action
\begin{equation}\label{1}
S_F=\frac{i}{2\pi}\int d^{\,2}\sigma\bar{\Psi}^\mu\rho^a\partial_a\Psi^\mu+S[\beta,\gamma]
\end{equation}
the partition function can be achieved by evaluating the path integral
\begin{equation}\label{2}
Z_F=\int D\Psi^{\mu}D\beta D\gamma e^{iS[\Psi,\bar{\Psi}]+iS[\beta,\gamma]}
\end{equation}
Here the action $S[\beta,\gamma]$ stands for the superconformal ghosts action. We skip the explicit derivation of the contribution arising from the integration over the superconformal ghost fields $\beta$ and $\gamma$ as its net effect is to decrease the space-time dimensions by 2. After some algebra and upon introducing $\Psi^{\mu }=(\psi^\mu,\tilde{\psi}^{\mu})^T$ the action (1) (with Euclidean signature for world-sheet and target space manifolds) takes a more simple form
\begin{equation}\label{3}
S[\psi,\tilde\psi]=\frac{i}{2\pi}\int d^2\sigma(\psi^\mu\bar\partial\psi^\mu
+\tilde{\psi}^\mu\partial\tilde{\psi}^\mu)
\end{equation}
where $\partial=\partial_\tau+i \partial_\sigma$. Now, we introduce the notation $(\pm,\pm)$ to distinguish the four fermionic spin structures in such a way that the upper (lower) sign denotes the periodic (anti-periodic) boundary conditions along the $\tau$ and $\sigma$ directions, respectively [13, 14]. Therefore, the spin structures $(\pm,+)$ arise from the Ramond sector, which for the off-shell fluctuations implies
\begin{equation}\label{4}
\psi^\mu(\tau,\sigma)=\sum_{m\in\mathbb Z+\frac{1}{2},\,n\in\mathbb Z}\frac{\psi_{mn}^\mu}{\sqrt 2} u_{mn},
\qquad\psi^\mu(\tau,\sigma)=\sum_{m\in\mathbb Z,\,n\in\mathbb Z}\frac{\psi_{mn}^\mu}{\sqrt 2} u_{mn}
\end{equation}
with the eigen-mode $u_{mn}=e^{i\tau\omega_m}e^{-in\sigma}$. For the spin structures $(\pm,-)$, arising from the Neveu-Schwrtz sector, we have
\begin{equation}\label{5}
\psi^\mu(\tau,\sigma)=\sum_{m\in\mathbb Z,\,n\in\mathbb Z+\frac{1}{2}}\frac{\psi_{mn}^\mu}{\sqrt 2}u_{mn},\qquad\psi^\mu(\tau,\sigma)=\sum_{m,\,n\in\mathbb Z+\frac{1}{2}}\frac{\psi_{mn}^\mu}{\sqrt 2}u_{mn}
\end{equation}
In a similar way the fourier expansions of $\tilde{\psi}^\mu$ associated with different spin structures can be achieved via the substitution $u_{mn}(\tau,\sigma)\rightarrow u_{mn}(\tau,-\sigma)\equiv\tilde u_{mn}(\tau,\sigma)$ in equations (4) and (5).  The eigen-modes fulfill the orthogonality relation
\begin{equation}\label{6}
\langle u_{mn}u_{m'n'}+ \widetilde{u}_{mn}
\widetilde{u}_{m'n'}\rangle=2\pi s\delta_{m+m'}\delta_{n+n'}
\end{equation}
Here we have defined
\begin{equation}\label{2}
\langle Q\rangle=\int_0^sd\tau\int_0^{\pi}d\sigma\ Q
\end{equation}
Hence, by taking into account the Grassmannian nature of coefficients, i.e. $\{\psi^\mu_{mn},\psi^\nu_{m^\prime n^\prime}\}=0$  one obtains the partition function of the open superstring
{\setlength\arraycolsep{2pt}
\begin{eqnarray}\label{7}
Z_{\psi}=\int D\psi^\mu e^{-S[\psi^\mu,\tilde{\psi}^\mu]}
=\prod_{mn}\lambda_{mn}^{\frac{d}{2}}
\end{eqnarray}}
with $\lambda_{mn}=-\frac{s}{2}(\omega_m+in)$. The above infinite product can be easily calculated with the aid of identities
{\setlength\arraycolsep{2pt}
\begin{eqnarray}\label{8}
\prod_{m\in\mathbb Z}(mx+y)=2\sinh\bigg(\frac{i\pi y}{x}\bigg),\qquad
\prod_{m\in\mathbb Z+\frac{1}{2}}(mx+y)=2\cosh\bigg(\frac{i\pi y}{x}\bigg)
\end{eqnarray}
and the zeta-function regularizations
{\setlength\arraycolsep{2pt}
\begin{eqnarray}\label{9}
\sum_{m\in\mathbb N}m=\frac{1}{12},\qquad\sum_{m\in\mathbb N-\frac{1}{2}}m=\frac{1}{24}
\end{eqnarray}
Therefore we find the open superstring partition function as [9] ($q=e^{-\frac{s}{2}}$)
\begin{equation}\label{10}
Z_{\psi}=\prod_{mn}\lambda_{mn}^{\frac{d}{2}}=\left\{\begin{array}{ll}
2^{\frac{d}{2}}\,q^{\frac{d}{12}}\prod_{n\in\mathbb N}(1+q^{2n})^d,\,\quad m\in\mathbb Z+\frac{1}{2}\\
q^{-\frac{d}{24}}\prod_{n\in\mathbb N}(1+q^{2n-1})^d,\quad m\in\mathbb Z+\frac{1}{2}\\
q^{-\frac{d}{24}}\prod_{n\in\mathbb N}(1-q^{2n-1})^d,\quad m\in\mathbb Z
\end{array}\right.
\end{equation}
We shall assign the symbols $Z^{-+}_\psi$, $Z^{--}_\psi$ and $Z^{+-}_\psi$ to the terms of equation (11) from above to below, respectively. One must note that $Z^{++}_\psi=0$ because of the well-known property of the Grassmann variables
\begin{equation}\label{11}
\int d\psi^{\mu}_{mn}=0
\end{equation}
\section*{\large 3\quad Fermionic Partition Function: The Case of Angled D1-Branes}
We specify the position of first D1-brane by
\begin{equation}\label{12}
X^i(\tau,0)=0,\quad\quad i=2,...,d
\end{equation}
and the second one by
{\setlength\arraycolsep{2pt}
\begin{eqnarray}\label{13}
X^2(\tau,\pi)\cos\alpha&=&X^1(\tau,\pi)\sin\alpha\\\nonumber
X^r(\tau,\pi)&=&l_r
\end{eqnarray}}
where $r=3,...,d$. We denote the deflection angle by $\alpha=\pi a$ and $0\leq a\leq 1$. Then, the conditions satisfied by the ends of an open string at the boundaries, imposed by the classical equations of motion, read [1, 10-12]
{\setlength\arraycolsep{2pt}
\begin{eqnarray}\label{14}
\partial_\sigma X^1(0,\tau)&=&
X^2(\tau,0)=0,\\\nonumber
\partial_\sigma X^1(\tau,\pi,)\cos\alpha&=&-\partial_\sigma X^2(\tau,\pi)\sin\alpha
\end{eqnarray}}
Similarly, for the fermionic degrees of freedom we find
\begin{equation}\label{15}
\bar{\epsilon}\rho^1\rho^0\psi^1(0,\tau)=\bar{\epsilon}\rho^1\rho^1\psi^1(0,\tau)=0
\end{equation}
and
\begin{equation}\label{16}
\bar{\epsilon}\rho^1\rho^0(\psi^1+\tan\alpha\psi^2)=
\bar{\epsilon}\rho^1\rho^1(\psi^2+\tan\alpha\psi^1)=0
\end{equation}
at the other end $\sigma=\pi$. So, for the classical solutions one finds [10]
\begin{equation}\label{17}
\left(\begin{array}{ccc}
\psi^1\\
\tilde\psi^1\\
\end{array}\right)=\frac{1}{2i}\sum_n\psi_n
\left(\begin{array}{ccc}
e^{in_a(\tau-\sigma)}\\
\pm e^{in_a(\tau+\sigma)}
\end{array}\right)+\textrm{complex conjugate}
\end{equation}
and
\begin{equation}\label{17}
\left(\begin{array}{ccc}
\psi^2\\
\tilde\psi^2\\
\end{array}\right)=\frac{1}{2}\sum_n\psi_n
\left(\begin{array}{ccc}
e^{in_a(\tau-\sigma)}\\
\mp e^{in_a(\tau+\sigma)}
\end{array}\right)+\textrm{complex conjugate}
\end{equation}
The lower sign in expression (18) and (19) corresponds to the NS sector. Now let us consider the fluctuations around the classical solutions in both sectors as
{\setlength\arraycolsep{2pt}
\begin{eqnarray}\label{20}
\psi^1&=&\frac{1}{2i}\sum_{m,n\in\mathbb Z}(\psi_{mn}u_{mn}^a-\bar\psi_{mn}\bar u_{mn}^a)\\
\psi^2&=&\frac{1}{2}\sum_{m,n\in\mathbb Z}(\psi_{mn}u_{mn}^a+\bar\psi_{mn}\bar u_{mn}^a)
\end{eqnarray}}
and
{\setlength\arraycolsep{2pt}
\begin{eqnarray}\label{20}
\tilde\psi^1&=&\frac{1}{2i}\sum_{m,n \in\mathbb Z}(\psi_{mn}\tilde u_{mn}^a-\bar\psi_{mn}\bar{\tilde u}_{mn}^a)\\
\tilde\psi^2&=&\frac{1}{2}\sum_{m,n \in\mathbb Z}(\psi_{mn}\tilde u_{mn}^a+\bar\psi_{mn}\bar{\tilde u}_{mn}^a)
\end{eqnarray}}
where $u_{mn}^a=e^{i\omega_m\tau}e^{-in_a\sigma}$ and $n_a=n+a$. Thus one finds
\begin{equation}\label{17}
\sum_{A=1}^2\langle\psi^A\bar\partial\psi^A\rangle=\frac{1}{4}
\sum_{mn}\sum_{m'n'}\Psi^{\textrm{t}}_{mn}\left(\begin{array}{ccc}
0&-2l_{m'n'}^a\langle u_{mn}^a\bar u_{m'n'}^a\rangle\\
2l_{m'n'}^a\langle\bar u_{mn}^a u_{m'n'}^a\rangle&0
\end{array}\right)\Psi_{m'n'}
\end{equation}
where we have introduced $\Psi_{mn}^{\textrm{t}}=(\psi_{mn},\bar\psi_{mn})$ and  $l^a_{mn}=(i\omega_m-n_a)$. This expression when combined with $\sum_{A=1}^2\langle\tilde\psi^A\partial\tilde\psi^A\rangle$ yields the diagonalized action
\begin{equation}\label{17}
S=
\sum_{mn}\Psi^{\textrm{t}}_{mn}\left(\begin{array}{ccc}
0&-\lambda_{mn}^a\\
\lambda_{mn}^a&0
\end{array}\right)\Psi_{mn}
\end{equation}
where we have gained
\begin{equation}\label{1}
\langle\bar u^a_{mn}u^a_{m'n'}+\bar {\tilde u}^a_{mn}\tilde u^a_{m'n'}\rangle=2\pi s\delta_{n,n'}\delta_{m,m'}
\end{equation}
Thus integration over the fields $\psi^1$ and $\psi^2$, or equivalently over $\Psi$, yields
\begin{equation}\label{22}
Z_{\psi^{1}\psi^{2}}=\int D\Psi e^{-S[\Psi]}=\prod_{mn}\det\left(\begin{array}{ccc}
0&-\lambda_{mn}^a\\
\lambda_{mn}^a&0
\end{array}\right)^{\frac{1}{2}}=\prod_{mn}\lambda_{mn}^a
\end{equation}
 Therefore, on invoking the well-known formula
\begin{equation}\label{23}
\sum^\infty_{n=1}(n+a)=\frac{1}{24}-\frac{1}{2}\bigg(a+\frac{1}{2}\bigg)^2
\end{equation}
we obtain
{\setlength\arraycolsep{2pt}
\begin{eqnarray}\label{20}
Z_{\psi^{1}\psi^{2}}
&=&\left\{\begin{array}{ll}
q^{\frac{1}{6}+a(a-1)}(1+q^{2a})\prod_{n\in\mathbb N}(1+q^{2n+2a})(1+q^{2n-2a}),\,\;\quad m\in\mathbb Z+\frac{1}{2}\\
q^{-\frac{1}{12}+a^2}\prod_{n\in\mathbb N}(1+q^{2n+2a-1})(1+q^{2n-2a-1}),\,\qquad\qquad m\in\mathbb Z+\frac{1}{2}\\
q^{-\frac{1}{12}+a^2}\prod_{n\in\mathbb N}(1-q^{2n+2a-1})(1-q^{2n-2a-1}),\,\qquad\qquad m\in\mathbb Z
\end{array}\right.
\end{eqnarray}}
The remaining bosonic degrees of freedom either satisfy the Neumann or the Dirichlet boundary condition. So, the corresponding fermionic degrees of freedom are characterized by the condition
\begin{equation}\label{25}
\psi(\tau,\pi)=\pm\eta\tilde\psi(\tau,\pi)
\end{equation}
with $\eta=1$ ($\eta=-1$) for the Neumann (Dirichlet) boundary condition. The minus sign refers to the Dirichlet boundary condition [15]. However, for a typical fermionic degree of freedom the partition function regardless to the boundary condition satisfied by its bosonic counterpart is given by equation (11) (with $d=1$). So, by taking into account the invariance under the modular transformation, we find the fermionic partition function $Z_F=Z_\psi Z_{\beta\gamma}$ in $d=10$ dimensions as
\begin{equation}\label{26}
Z_F=\frac{1}{2}\bigg[-Z^{+-}_{\psi^{1}\psi^{2}}(Z^{+-}_{\psi})^6+Z^{--}_{\psi^{1}\psi^{2}}(Z^{--}_{\psi})^6+Z^{-+}
_{\psi^{1}\psi^{2}}(Z^{-+}_{\psi})^6\bigg]
\end{equation}
where the factor $\frac{1}{2}$ comes from the GSO projection. A similar result for fermionic partition function is derived earlier by applying the harmonic oscillator representation [10-12].
The one loop vacuum energy of the system becomes
\begin{equation}\label{27}
\mathcal{A}=\ln\mathcal Z=\int^{\infty}_{0}\frac{ds}{s}\mathcal Z(s)
\end{equation}
where the superstring partition function in non-parallel D1-brane setup will be
{\setlength\arraycolsep{2pt}
\begin{eqnarray}\label{28}
\mathcal Z&=&Z_FZ_B\\\nonumber
&=&Tq^{\frac{Y^2}{\pi^2}}(1-q^{2a})^{-1}\prod_{n\in\mathbb N}(1-q^{2n})^{-6}(1-q^{2n+2a})^{-1}(1-q^{2n-2a})^{-1}\\\nonumber
&\times&\frac{1}{2}\bigg[-8(1+q^{2a})\prod_{n=1}(1+q^n)^6(1+q^{2n+2a})(1+q^{2n-2a})\\\nonumber
&+&q^{-1+a}\prod_{n\in\mathbb N}(1+q^{2n-1})^6(1+q^{2n+2a-1})(1+q^{2n-2a-1})\\\nonumber
&+&q^{-1+a}\prod_{n\in\mathbb N}(1-q^{2n-1})^6(1-q^{2n+2a-1})(1-q^{2n-2a-1})\bigg]
\end{eqnarray}}
Here the partition function of bosonic part (in 26 dimensions) is [1, 10-12]
\begin{equation}\label{29}
Z_B=\frac{T}{\sqrt{2\pi s}}\frac{q^{\frac{Y^2}{\pi^2}+a(a-1)-2}}{1-q^{2a}}\prod_{n\in\mathbb N}(1-q^{2n})^{-22}(1-q^{2n+2a})^{-1}(1-q^{2n-2a})^{-1}
\end{equation}
where $Y^2$ stands for the distance between D-branes. The total interval of interaction time $T$ arises from integration over the zero mode of $X^0$.
\section*{\large References}
[1] \hspace{0.2cm}A. Jahan, Mod. Phys. Lett. \textbf {A25} (2010) 619.\\\
[2] \hspace{0.2cm}A. M. Polyakov, Phys. Lett. \textbf{B103} (1981) 207.\\\
[3] \hspace{0.2cm}A. M. Ployakov, Phys. Lett. \textbf{B103} (1981) 211.\\\
[4] \hspace{0.2cm}C. P. Burgess, Nucl. Phys. \textbf{B291} (1987) 256.\\\
[5] \hspace{0.2cm}C. P. Burgess, Nucl. Phys. \textbf{B291} (1987) 285.\\\
[6] \hspace{0.2cm}J. Polchinski, Commun. Math. Phys. \textbf{104} (1986) 37.\\\
[7] \hspace{0.2cm}C. S. Howe, B. Sakita, and M. A . Virasoro, Phys. Rev. \textbf{D2} (1970) 2857.\\\
[8] \hspace{0.2cm}D. B. Fairlie and H.B. Nielsen, Nucl. Phys.\textbf{ B20} (1970) 637.\\\
[9] \hspace{0.2cm}P. Ginsparg, Les Houches, Session XLIX, 1988, \textit{Fields, Strings and
Critical Phenomena}, ed. by E. Brezin and J. Zinn-Justin,
Elsevier Science Publishers B.V. (1989).\\\
[10] \hspace{0.2cm}A. Matusis, Int. J. Mod. Phys. \textbf{A14} (1999) 1153.\\\
[11] \hspace{0.2cm}H. Arfaei, M. M. Sheikh-Jabbari, Phys. Lett. \textbf{B394} (1997) 288.\\\
[12] \hspace{0.2cm}T. Kiato, N. Ohta and J. Zhaou, Phys. Lett. \textbf{B428} (1998) 68.\\\
[13] \hspace{0.2cm}J. Ambjorn, Y. M. Makeenko, G. W. Semenoff and R. J. Szabo, JHEP \textbf{0302} (2003) 026. \\\
[14] \hspace{0.2cm}M. Green, J. Schwartz, E. Witten, \textit{Superstring Tehory}, Vol. 1, Cambridge University Press, 1987.\\\
[15] \hspace{0.2cm}P. Di Vecchia, A. Liccardo, \textit{D-Branes in String Theory I}, NATO Adv. Study Inst. Ser. C. Math. Phys. Sci. 556 (2000).
\end{document}